\documentclass[amssymb,aps,pre,showpacs,manuscript]{revtex4}
\pdfoutput=1
\usepackage{mathrsfs}
\usepackage{amsmath}
\usepackage{amssymb}
\usepackage{graphicx}
\usepackage{esint}

\begin{document}

\title{Variational Superposed Gaussian Approximation for \\
 Time-dependent Solutions of Langevin Equations}

\author{Yoshihiko Hasegawa}

\email[Corresponding author~: ]{yoshihiko.hasegawa@gmail.com}

\date{\today}

\affiliation{Department of Information and Communication Engineering, Graduate
School of Information Science and Technology, The University of Tokyo,
Tokyo 113-8656, Japan}
\begin{abstract}
We propose a variational superposed Gaussian approximation (VSGA)
for dynamical solutions of Langevin equations subject to applied signals,
determining time-dependent parameters of superposed Gaussian distributions
by the variational principle. We apply the proposed VSGA to systems
driven by a chaotic signal, where the conventional Fourier method
cannot be adopted, and calculate the time evolution of probability
density functions (PDFs) and moments. Both white and colored Gaussian
noises terms are included to describe fluctuations. Our calculations
show that time-dependent PDFs obtained by VSGA agree excellently with
those obtained by Monte Carlo simulations. The correlation between
the chaotic input signal and the mean response are also calculated
as a function of the noise intensity, which confirms the occurrence
of aperiodic stochastic resonance with both white and colored noises.

\end{abstract}


\maketitle

\section{Introduction}

Langevin equations can model systems subject to fluctuations and hence
have many applications in diverse research fields such as physics,
chemistry, financial engineering, and biology \cite{Kampen:1992:StocProcBook,Hasegawa:2013:OptimalPRC,Hasegawa:2014:PRL}.
Without a driving force, systems subject to white Gaussian noise relax
to their stationary states. For one-dimensional stationary systems,
the probability density function (PDF) can be obtained in a closed
form for many cases. However, in the presence of a driving force,
its time-dependent solution is rarely available even for one-dimensional
systems. Recent advancements in nonequilibrium theory \cite{Ritort:NEArticle,Seifert:2012:FTReview}
strongly demand reliable methods for time-dependent solutions of Langevin
equations for systems driven by time-dependent external forces. The
moment method (MM) is widely used to study dynamics \cite{Huang:1995:MM4FPE,Rodriguez:1996:SpikeNeurons,Tuckwell:2009:HHeq};
it considers the time evolution of moments of PDFs {[}in most cases,
up to the second-order moments (mean and variance) are considered{]}.
If we truncate at the second moment (i.e., $n$th-order terms where
$n\ge3$ are ignored), the number of differential equations is $N(N+3)/2$,
where $N$ is the dimensionality of the model; thus, with current
computer capabilities, the MM is tractable up to relatively large
$N$. Although the MM can provide satisfactory results for linear
(or weakly nonlinear) systems, its applicability collapses even for
simple bistable models. Here, for time-dependent solutions of Langevin
equations, we propose an approximation technique in which PDFs are
represented by superposed multiple Gaussian distributions, obtaining
time-evolution equations for parameters of each of the Gaussian distributions
with the variational principle. We call the proposed method the variational
superposed Gaussian approximation (VSGA). Dynamical Gaussian approximations
have a long history in quantum mechanics. Heller introduced the Gaussian
wavepacket method \cite{Heller:1976:TDV}, which approximates time-dependent
solutions of Schr\"odinger equations with a Gaussian packet by obtaining
equations for the mean and variance through the McLachlan variational
principle \cite{McLachlan:1964:VP} (other equivalent variational
principles are also known \cite{Dirac:1930:VP,Frenkel:1934:WaveMech,Broeckhove:1988:VP}
and this is a special case of the weighted residual method). Several
researchers extended Heller's approach to incorporate multiple Gaussian
distributions \cite{Skodje:1984:MultiPackets,Worth:2004:MultiGWP,Zoppe:2005:GPackets},
and these methods can provide reliable solutions for time-dependent
wave functions by virtue of their multiplicity. Although the effectiveness
of the multiple Gaussian method with the variational principle has
been shown to approximate time-dependent wave functions \cite{Skodje:1984:MultiPackets,Worth:2004:MultiGWP,Zoppe:2005:GPackets},
its capability has not been shown in the context of time-dependent
Fokker-Planck equations (FPEs).

After the Gaussian wavepacket approximations in quantum mechanics,
several studies employed a superposition of Gaussian distributions
for Langevin equations \cite{Er:1998:MultiGaussian,Pradlwarter:2001:NonLinear,Terejanu:2008:GaussMixture}.
Reference~\cite{Er:1998:MultiGaussian} adopted superposed Gaussian
distributions to approximate stationary solutions through the weighted
residual method. For dynamical solutions, Ref.~\cite{Pradlwarter:2001:NonLinear}
employed superposed Gaussian distributions based on the statistical
equivalent linearization where PDFs are represented by small elements
of Gaussian distributions. It was noted that Pradlwarter's method
has to manage the variance and the number of Gaussian distributions
during the propagation. Similarly, Terejanu \emph{et al.} \cite{Terejanu:2008:GaussMixture}
developed an approximation scheme based on superposed Gaussian distributions,
which calculated the mean and variance with a fixed weight. After
calculating the mean and variance, they optimized the weight, using
quadratic programming to minimize the squared error. Unlike these
approaches, the VSGA does not require such extra steps; it directly
calculates the mean, variance and weight in a unified way.  Reference~\cite{Paola:2002:ApproximateFPE}
approximates time-dependent solutions with exponential of a polynomial
function to obtain the time-evolution equations of parameters through
the weighted residual method. However, such an approximation has difficulty
in satisfying the normalization condition during the time evolution.
There are several numerical approaches to the study of the dynamics
of FPEs that represent PDFs by using complete set functions (e.g.,
a matrix continued-fraction method; for details, see Ref.~\cite{Risken:1989:FPEBook}
and the references therein). More-direct numerical schemes, such as
a finite-element method \cite{Harrison:1988:FEM4FPE,Kumar:2006:FPEbyFEM}
and a finite-difference method \cite{Whitney:1970:FDM4FPE}, have
also been studied. These approaches, however, have high computational
costs and hence are not suited for time-dependent solutions. 

To investigate the effectiveness of VSGA, we applied it to a quartic
bistable system subject to white or colored noise. Although a bistable
system can describe switching dynamics and has many and varied applications
to realistic problems \cite{Wilhelm:2009:Bistability}, its nonlinearity
makes the application of the simple MM difficult. We consider a system
driven by a chaotic signal (the R\"ossler oscillator). When the signal
is aperiodic, we cannot use a Fourier series expansion, as is often
employed for periodic cases \cite{Jung:1993:PeriodicSystem}, and
hence many studies have resorted to using direct Monte Carlo (MC)
simulations. We show that VSGA can accurately approximate the time-dependent
moments and the PDFs of the systems for both white (one-dimensional)
and colored (two-dimensional) noises. Calculating the correlation
between the chaotic input signal and the mean of the dynamics {[}cf.
Eq.~\eqref{eq:ASR_def}{]}, we show that the correlation is maximal
when the noise is of intermediate strength; this is a signature of
aperiodic stochastic resonance (ASR) \cite{Collins:1995:ASRinExcite,Collins:1996:AperiodicSR}
{[}for general stochastic resonance (SR), see \cite{Benzi:1981:SR,McNamara:1989:SR,Gammaitoni:1998:SR,McDonnell:2008:SRBook,McDonnell:2009:SR}
and the references therein{]}. Furthermore, from the results with
colored noise, we show that the time correlation weakens the magnitude
of the ASR.

The remainder of this paper is organized as follows. In Section~\ref{sec:methods},
we introduce our proposed method, the VSGA, and provide a detailed
explanation of the variational principle for FPEs. We obtain implicit
differential equations that should be satisfied by the mean, variance,
and weights of each of the Gaussian distributions. In Section~\ref{sec:results},
we investigate the effectiveness of VSGA by applying it to two cases:
a bistable system driven by a chaotic signal subject to white noise
(Section~\ref{sub:one_dim}), and colored noise (Section~\ref{sub:two_dim}).
Finally, we provide a discussion and present our conclusions in Section~\ref{sec:discussion}.

\section{Methods\label{sec:methods}}

\begin{figure}
\includegraphics[width=12cm]{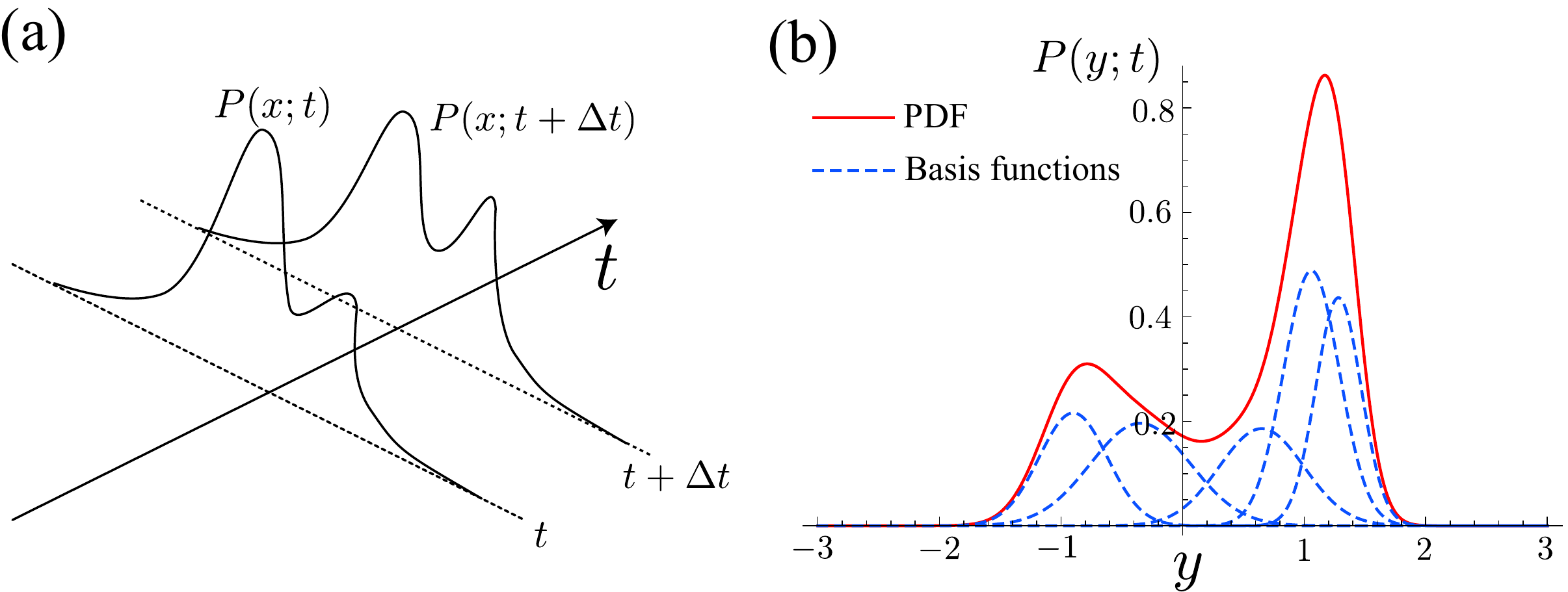}

\caption{(Color online) (a) Illustration of time evolution of a PDF. If the
PDF at time $t$ {[}i.e., $P(x;t)${]} is known, the optimal $P(x;t+\Delta t)$
is given by $P(x;t+\Delta)\simeq P(x;t)+\Delta t\varTheta(x;t)$,
where $\varTheta(x;t)$ is optimal. (b) Example of an approximative
PDF (solid line) that is a superposition of five Gaussian distributions
(dashed lines). \label{fig:time_evolution}}
\end{figure}

We consider a $N$-dimensional Langevin equation (the Stratonovich
interpretation) 
\begin{equation}
\frac{dx_{i}}{dt}=f_{i}(\boldsymbol{x},t)+\sum_{j=1}^{N_{g}}g_{ij}(\boldsymbol{x},t)\xi_{j}(t),\hspace{1em}(i=1,2,..,N),\label{eq:Langevin}
\end{equation}
where $\boldsymbol{x}=(x_{1},...,x_{N})^{\top}$ ($\top$ denotes
the transpose operation) is an $N$-dimensional column vector, $f_{i}(\boldsymbol{x},t)$
and $g_{ij}(\boldsymbol{x},t)$ denote drift and multiplicative terms,
respectively, $\xi_{i}(t)$ is white Gaussian noise with the correlation
$\left\langle \xi_{i}(t)\xi_{j}(t^{\prime})\right\rangle =2\delta_{ij}\delta(t-t^{\prime})$,
and $N_{g}$ is the number of noise sources \cite{Risken:1989:FPEBook}.
The Langevin equation \eqref{eq:Langevin} has the corresponding FPE
\cite{Risken:1989:FPEBook} 
\begin{equation}
\frac{\partial}{\partial t}P(\boldsymbol{x};t)=\hat{L}(\boldsymbol{x},t)P(\boldsymbol{x};t),\label{eq:FPE}
\end{equation}
where $P(\boldsymbol{x};t)$ is the probability density of $\boldsymbol{x}$
at time $t$, and $\hat{L}(\boldsymbol{x},t)$ is an FPE operator
defined by 
\begin{equation}
\hat{L}(\boldsymbol{x},t)=-\sum_{i}\frac{\partial}{\partial x_{i}}F_{i}(\boldsymbol{x},t)+\sum_{i,j}\frac{\partial^{2}}{\partial x_{i}\partial x_{j}}G_{ij}(\boldsymbol{x},t).\label{eq:L_def}
\end{equation}
Here $F_{i}(\boldsymbol{x},t)=f_{i}(\boldsymbol{x},t)+\sum_{k,j}g_{kj}(\boldsymbol{x},t)\partial_{x_{k}}g_{ij}(\boldsymbol{x},t)$
and $G_{ij}(\boldsymbol{x},t)=\sum_{k}g_{ik}(\boldsymbol{x},t)g_{jk}(\boldsymbol{x},t)$
(note that the VSGA can be applied to the It\^o interpretation by
modifying $F_{i}(\boldsymbol{x},t)$) \cite{Risken:1989:FPEBook}.
We are interested in a time-dependent solution $P(\boldsymbol{x};t)$
of Eq.~\eqref{eq:FPE}. We approximate the time evolution by using
the variational principle, which is explained below for the FPE.

Let $\varTheta(\boldsymbol{x};t)$ be the time derivative of $P(\boldsymbol{x};t)$,
i.e., $\varTheta(\boldsymbol{x};t)=\dot{P}(\boldsymbol{x};t)$. We
focus on a specific time $t$, where $P(\boldsymbol{x};t)$ is already
known, and we want to know the optimal time evolution $\varTheta(\boldsymbol{x};t)$
{[}Fig.~\ref{fig:time_evolution}(a){]}. In other words, we want
to calculate $P(\boldsymbol{x};t+\Delta t)$, where $\Delta t$ is
a sufficiently small increment, from a known $P(\boldsymbol{x};t)$
by using $P(\boldsymbol{x};t+\Delta t)\simeq P(\boldsymbol{x};t)+\Delta t\varTheta(\boldsymbol{x};t)$.
From Eq.~\eqref{eq:FPE}, the optimal $\varTheta(\boldsymbol{x};t)$
should minimize 
\begin{equation}
R[\varTheta]=\int_{-\infty}^{\infty}\left\{ \hat{L}(\boldsymbol{x},t)P(\boldsymbol{x};t)-\varTheta(\boldsymbol{x};t)\right\} ^{2}d\boldsymbol{x},\label{eq:residue}
\end{equation}
where we have abbreviated as follows: $\int_{-\infty}^{\infty}dx_{1}\cdots\int_{-\infty}^{\infty}dx_{N}=\int_{-\infty}^{\infty}d\boldsymbol{x}$.
Although we may obtain the optimal $P(\boldsymbol{x};t+\Delta t)$
by solving Eq.~\eqref{eq:residue} with respect to $\varTheta$,
the optimal $\varTheta$ does not necessarily yield solutions that
satisfy the normalization condition $\int_{-\infty}^{\infty}P(\boldsymbol{x};t)d\boldsymbol{x}=1$
at any time $t$. Therefore we should impose an additional constraint
on Eq.~\eqref{eq:residue}. When $P(\boldsymbol{x};t)$ is normalized
at $t=0$, then the normalization of $P(\boldsymbol{x};t)$ at $t>0$
is satisfied by the equation given by 
\begin{align}
\frac{d}{dt}\int_{-\infty}^{\infty}P(\boldsymbol{x};t)d\boldsymbol{x} & =\int_{-\infty}^{\infty}\varTheta(\boldsymbol{x};t)d\boldsymbol{x}=0.\label{eq:constraint}
\end{align}
Therefore, to minimize Eq.~\eqref{eq:residue} with the normalization
condition of Eq.~\eqref{eq:constraint}, we consider the following
equation: 
\begin{equation}
\tilde{R}[\varTheta]=\int_{-\infty}^{\infty}\left\{ \hat{L}(\boldsymbol{x},t)P(\boldsymbol{x};t)-\varTheta(\boldsymbol{x};t)\right\} ^{2}d\boldsymbol{x}+\lambda\int_{-\infty}^{\infty}\varTheta(\boldsymbol{x};t)d\boldsymbol{x},\label{eq:tilde_R}
\end{equation}
where $\lambda$ is the Lagrange multiplier. With a variation of $\delta\varTheta(\boldsymbol{x};t)$
in Eq.~\eqref{eq:tilde_R}, $\delta\tilde{R}$ should vanish for
the optimal $\varTheta(\boldsymbol{x};t)$, yielding 
\begin{equation}
\int_{-\infty}^{\infty}\delta\varTheta\left\{ \hat{L}(\boldsymbol{x},t)P(\boldsymbol{x};t)-\varTheta(\boldsymbol{x};t)+\lambda\right\} d\boldsymbol{x}=0,\label{eq:MVP_FPE}
\end{equation}
where we redefined $\lambda$ for notational convenience. Suppose
$P(\boldsymbol{x};t)$ is a function parametrized by time-dependent
$K$ values $\boldsymbol{\theta}(t)=(\theta_{1}(t),\theta_{2}(t),..,\theta_{K}(t))$:
\begin{equation}
P(\boldsymbol{x};t)=P(\boldsymbol{x};\boldsymbol{\theta}(t)),\label{eq:Psi_param}
\end{equation}
where the time-dependence of $P(\boldsymbol{x};t)$ is represented
through $\boldsymbol{\theta}(t)$. Thus $\varTheta(\boldsymbol{x};t)$
is given by 
\begin{equation}
\varTheta(\boldsymbol{x};t)=\varTheta(\boldsymbol{x};\boldsymbol{\theta}(t),\dot{\boldsymbol{\theta}}(t)).\label{eq:Theta_param}
\end{equation}
The variation $\delta\varTheta$ can be achieved only through the
variation $\Delta\dot{\boldsymbol{\theta}}$ ($\boldsymbol{\theta}$
cannot be changed, as we assumed that $P(\boldsymbol{x};t)=P(\boldsymbol{x};\boldsymbol{\theta}(t))$
is fixed at time $t$): 
\begin{equation}
\delta\varTheta=\sum_{\ell=1}^{K}\frac{\partial\varTheta(\boldsymbol{x};\boldsymbol{\theta},\dot{\boldsymbol{\theta}})}{\partial\dot{\theta}_{\ell}}\Delta\dot{\theta}_{\ell},\label{eq:relation}
\end{equation}
where 
\begin{align}
\frac{\partial\varTheta(\boldsymbol{x};\boldsymbol{\theta},\dot{\boldsymbol{\theta}})}{\partial\dot{\theta}_{\ell}} & =\frac{\partial}{\partial\dot{\theta}_{\ell}}\frac{dP(\boldsymbol{x};\boldsymbol{\theta})}{dt},\nonumber \\
 & =\frac{\partial}{\partial\dot{\theta}_{\ell}}\left(\sum_{\ell^{\prime}=1}^{K}\frac{\partial P(\boldsymbol{x};\boldsymbol{\theta})}{\partial\theta_{\ell^{\prime}}}\dot{\theta}_{\ell^{\prime}}\right),\nonumber \\
 & =\frac{\partial P(\boldsymbol{x};\boldsymbol{\theta})}{\partial\theta_{\ell}}.\label{eq:derivative_calc}
\end{align}
For the variation $\Delta\dot{\boldsymbol{\theta}}=(\Delta\dot{\theta}_{1},...,\Delta\dot{\theta}_{K})$,
we consider the simplest orthogonal case: 
\[
\Delta\dot{\boldsymbol{\theta}}=(1,0,0,\cdots,0),(0,1,0,\cdots,0),\cdots,(0,0,0,\cdots,1).
\]
Substituting Eqs.~\eqref{eq:relation} and \eqref{eq:derivative_calc}
into Eq.~\eqref{eq:MVP_FPE}, we have $K$ constraints 
\begin{equation}
\int_{-\infty}^{\infty}\frac{\partial P(\boldsymbol{x};\boldsymbol{\theta})}{\partial\theta_{\ell}}\left\{ \hat{L}(\boldsymbol{x},t)P(\boldsymbol{x};\boldsymbol{\theta})-\varTheta(\boldsymbol{x};\boldsymbol{\theta},\dot{\boldsymbol{\theta}})+\lambda\right\} d\boldsymbol{x}=0\hspace{1em}(\ell=1,2,..,K),\label{eq:MVP_FPE2}
\end{equation}
which is the variational principle for FPEs that is equivalent to
the McLachlan one. Also Eq.~\eqref{eq:MVP_FPE2} without $\lambda$
can be seen as minimizing the residual with a weight function $\partial_{\theta_{\ell}}P(\boldsymbol{x};\boldsymbol{\theta})$. 

We next show an explicit form of $P(\boldsymbol{x};\boldsymbol{\theta}(t))$.
We approximate $P(\boldsymbol{x};t)$ with a superposition of multiple
Gaussian distributions (Fig.~\ref{fig:time_evolution}(b)): 
\begin{equation}
P(\boldsymbol{x};\boldsymbol{\theta}(t))=\sum_{m=1}^{N_{B}}r_{m}g(\boldsymbol{x};\boldsymbol{A}_{m}(t),\boldsymbol{b}_{m}(t)),\label{eq:MGMM_def}
\end{equation}
where $g(\boldsymbol{x};\boldsymbol{A},\boldsymbol{b})$ is an unnormalized
Gaussian distribution: 
\begin{equation}
g(\boldsymbol{x};\boldsymbol{A},\boldsymbol{b})=\exp\left(-\boldsymbol{x}^{\top}\boldsymbol{A}\boldsymbol{x}+\boldsymbol{b}^{\top}\boldsymbol{x}\right).\label{eq:Gaussian_def}
\end{equation}
Here $\boldsymbol{A}$ is an $N\times N$ symmetric matrix (positive
definite), $\boldsymbol{b}$ is an $N$-dimensional column vector,
$r_{m}$ is a parameter that combines the weight of the $m$th Gaussian
with a normalization constant, and $N_{B}$ is the number of basis
functions. We employed a parametrization of Eq.~\eqref{eq:Gaussian_def}
that is different from the conventional multivariate Gaussian representation,
because multidimensional calculations are easier with Eq.~\eqref{eq:Gaussian_def}
(cf. Appendix A). For instance, multiplication is simply given by
\[
g(\boldsymbol{x};\boldsymbol{A}_{m},\boldsymbol{b}_{m})g(\boldsymbol{x};\boldsymbol{A}_{m^{\prime}},\boldsymbol{b}_{m^{\prime}})=g(\boldsymbol{x};\boldsymbol{A}_{m}+\boldsymbol{A}_{m^{\prime}},\boldsymbol{b}_{m}+\boldsymbol{b}_{m^{\prime}}).
\]
We optimized all of the Gaussian parameters by using the variational
principle, i.e., $\boldsymbol{\theta}=(\boldsymbol{A}_{m},\boldsymbol{b}_{m},r_{m})_{m=1}^{N_{B}}$.
For an $N$-dimensional system and $N_{B}$ basis functions, the total
number of parameters is 
\begin{equation}
K=\frac{N_{B}(N+1)(N+2)}{2}.\label{eq:K_def}
\end{equation}
From Eq.~\eqref{eq:MVP_FPE2} and the constraint of Eq.~\eqref{eq:constraint},
we obtain $(K+1)$ implicit differential equations of the following
form: 
\begin{equation}
H_{\ell}(\boldsymbol{\theta}(t),\dot{\boldsymbol{\theta}}(t),\lambda(t),t)=0,\hspace{1em}\ell=1,2,\cdots,K,K+1.\label{eq:DAE}
\end{equation}
Equation~\eqref{eq:DAE} is called a differential algebraic equation
(DAE) \cite{Ascher:1998:DAE}. Because the dimensionality of $(\dot{\boldsymbol{\theta}}(t),\lambda(t))$
is $K+1$ and there are $K+1$ equations, we can uniquely specify
$(\dot{\boldsymbol{\theta}}(t),\lambda(t))$ given $\boldsymbol{\theta}(t)$.
However, because it is very difficult to explicitly solve Eq.~\eqref{eq:DAE}
with respect to $(\dot{\boldsymbol{\theta}}(t),\lambda(t))$ for higher
dimensional cases, we use a DAE solver in \emph{MATHEMATICA} 10 (NDSolve
function). VSGA does not accept arbitrary initial values, because
$(\boldsymbol{A}_{m},\boldsymbol{b}_{m},r_{m})_{m=1}^{N_{B}}$ should
satisfy the normalizing condition, which can be obtained from Eqs.~\eqref{eq:Ai_def}--\eqref{eq:ri_def}
(cf. Appendix~\ref{sec:init_DAE}).

\section{Results\label{sec:results}}

We applied the VSGA to two double-well systems driven by chaotic signals,
one subject to white Gaussian noise (Section~\ref{sub:one_dim})
and the other subject to colored Gaussian noise (Section~\ref{sub:two_dim}).
We also performed MC simulations to show the reliability of the VSGA.

\subsection{Chaotically-driven bistable potential subject to white noise\label{sub:one_dim}}

\begin{figure}
\includegraphics[width=12cm]{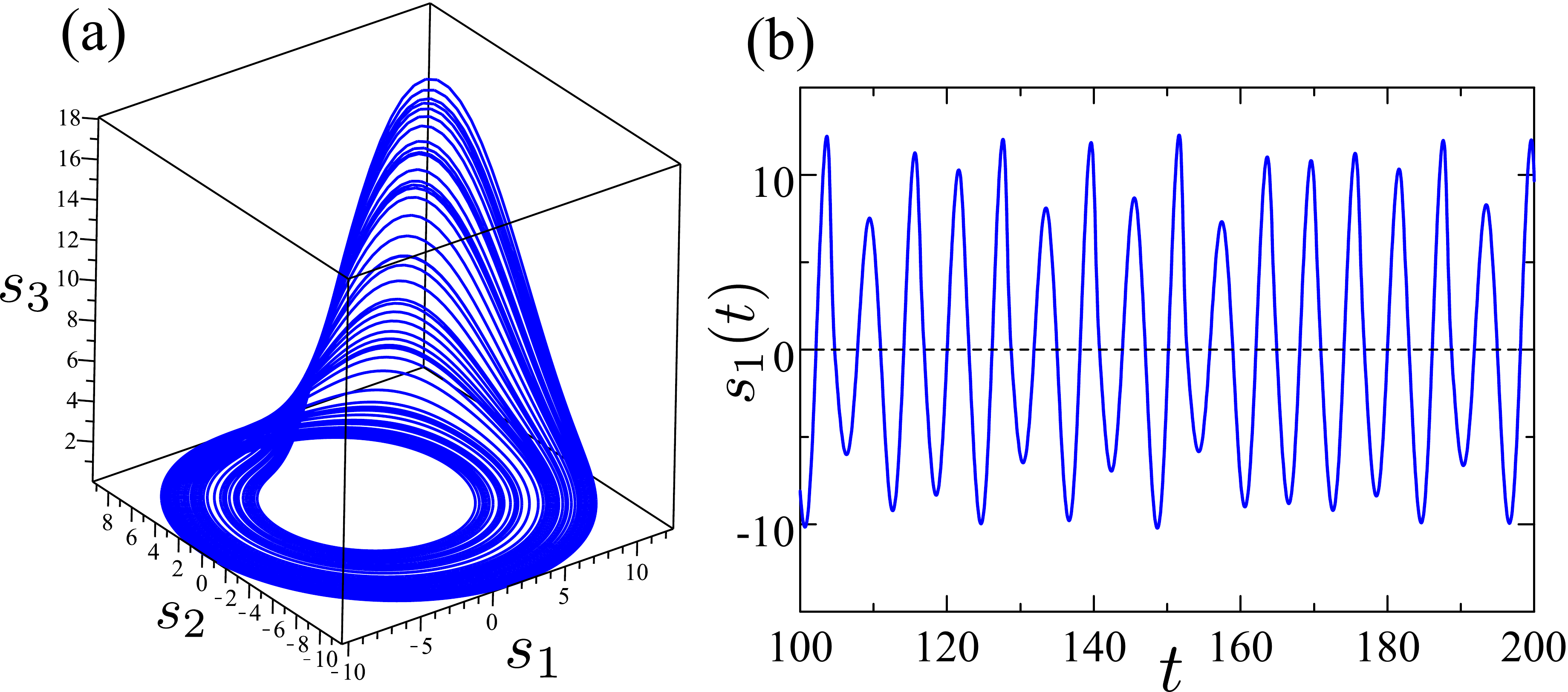}

\caption{(Color online) Trajectories of R\"ossler chaos {[}Eqs.~\eqref{eq:Rossler_1}--\eqref{eq:Rossler_3}{]}:
(a) 3D plot of $s_{1}$, $s_{2}$, and $s_{3}$; and (b) $s_{1}$
as a function of $t$. The parameters are $c_{1}=0.15$, $c_{2}=0.2$
and $c_{3}=7.1$. \label{fig:Rossler}}
\end{figure}

We applied the VSGA to a driven bistable potential subject to white
Gaussian noise. Specifically, we applied it to a one-dimensional potential
driven by an input signal $I(t)$: 
\begin{equation}
\frac{dy}{dt}=y-y^{3}+I(t)+\sqrt{D}\xi(t),\label{eq:Langevin_chaotic_bistable}
\end{equation}
where $D$ is the noise intensity and $\xi(t)$ is white Gaussian
noise with the correlation $\left\langle \xi(t)\xi(t^{\prime})\right\rangle =2\delta(t-t^{\prime})$.
The FPE operator $\hat{L}(\boldsymbol{x},t)=\hat{L}(y,t)$ is given
by 
\begin{equation}
\hat{L}(y,t)=-\frac{\partial}{\partial y}\left\{ y-y^{3}+I(t)\right\} +D\frac{\partial^{2}}{\partial y^{2}}.\label{eq:FPE_1d}
\end{equation}
Substituting Eq.~\eqref{eq:FPE_1d} into Eq.~\eqref{eq:MVP_FPE2},
we can calculate $K$ coupled DAEs with respect to $\boldsymbol{\theta}=(\boldsymbol{A}_{m},\boldsymbol{b}_{m},r_{m})_{m=1}^{N_{B}}$
(for the one-dimensional case, $\boldsymbol{A}_{m}=a_{m}$ and $\boldsymbol{b}_{m}=b_{m}$,
where $a_{m},b_{m}$ are real scalar quantities) requiring moments
of the Gaussian distribution of up to the sixth order ($\int_{-\infty}^{\infty}y^{n_{y}}g(y;\boldsymbol{A},\boldsymbol{b})dy$,
with $n_{y}\le6$). For the one-dimensional case, the normalization
constraint is 
\begin{align}
\frac{d}{dt}\int_{-\infty}^{\infty}P(\boldsymbol{x};\boldsymbol{\theta}(t))d\boldsymbol{x} & =\frac{d}{dt}\int_{-\infty}^{\infty}\sum_{m=1}^{N_{B}}r_{m}(t)\exp(-a_{m}(t)y^{2}+b_{m}(t)y)dy=0,\label{eq:constraint_1d}
\end{align}
yielding 
\begin{align}
0= & \sum_{m=1}^{N_{B}}\frac{\sqrt{\pi}}{4a_{m}(t)^{5/2}}\left[4a_{m}(t)^{2}\dot{r}_{m}(t)-2a_{m}(t)r_{m}(t)\left\{ \dot{a}_{m}(t)-b_{m}(t)\dot{b}_{m}(t)\right\} \right.\nonumber \\
 & \left.-b_{m}(t)^{2}\dot{a}_{m}(t)r_{m}(t)\right]\exp\left(\frac{b_{m}(t)^{2}}{4a_{m}(t)}\right).\label{eq:norm_1d}
\end{align}
Equation~\eqref{eq:norm_1d} should be solved along with the DAEs
obtained from Eq.~\eqref{eq:MVP_FPE2}; the total dimensionality
of the DAEs is $K+1$.

For the input signal $I(t)$, we used the R\"ossler oscillator \cite{Rossler:1976:Chaos}:
\begin{align}
\frac{ds_{1}}{dt} & =-s_{2}-s_{3},\label{eq:Rossler_1}\\
\frac{ds_{2}}{dt} & =s_{1}+c_{1}s_{2},\label{eq:Rossler_2}\\
\frac{ds_{3}}{dt} & =c_{2}+s_{3}(s_{1}-c_{3}).\label{eq:Rossler_3}
\end{align}
Here, $c_{i}$ are parameters of the oscillator, and we used $c_{1}=0.15$,
$c_{2}=0.2$, and $c_{3}=7.1$ (identical to the values used in Ref.~\cite{Silchenko:1999:ChaosSR}),
with which Eqs.~\eqref{eq:Rossler_1}--\eqref{eq:Rossler_3} exhibit
chaotic dynamics. Figures~\ref{fig:Rossler}(a) and (b) show trajectories
of the R\"ossler oscillator for (a) $s_{1}$, $s_{2}$, and $s_{3}$,
and (b) $s_{1}$ as a function of time $t$. The average peak-to-peak
interval (which corresponds to the period of the oscillations) of
$s_{1}(t)$ is about $6$. We define the input signal as 
\begin{equation}
I(t)=\alpha s_{1}(\omega t),\label{eq:input_definition}
\end{equation}
where $\alpha$ is the input strength and $\omega$ is the reciprocal
of the time-scale (this corresponds to the angular frequency of the
periodic oscillations). Although time-dependent solutions of periodically
driven systems are often represented as a Fourier series expansion
\cite{Jung:1993:PeriodicSystem}, such an expansion cannot be used
for a chaotically driven system.

\begin{figure}
\includegraphics[width=16cm]{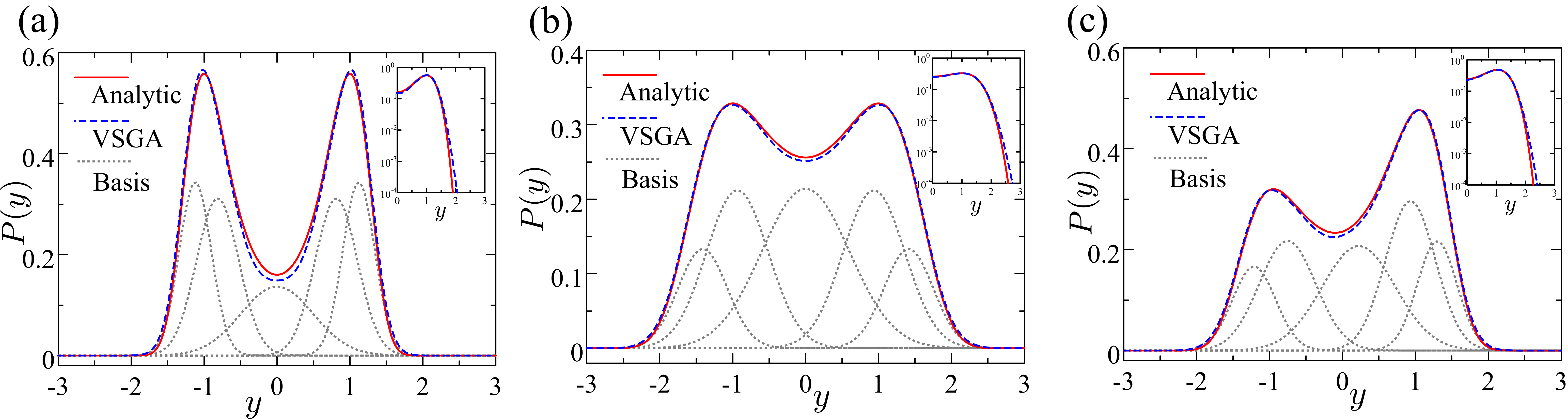}

\caption{(Color online) Stationary PDFs obtained by analytic calculation of
Eq.~\eqref{eq:statdist_1d} (solid line) and by the VSGA with $N_{B}=5$
(dashed line) for (a) $D=0.2$ and $\kappa=0$, (b) $D=1.0$ and $\kappa=0$,
and (c) $D=0.5$ and $\kappa=0.1$. In (a)--(c), the dotted lines
denote each of the single bases of the VSGA. The insets show log-plots
of PDFs at tail regions. \label{fig:statdist_1d}}
\end{figure}

\begin{figure}
\begin{centering}
\includegraphics[width=6cm]{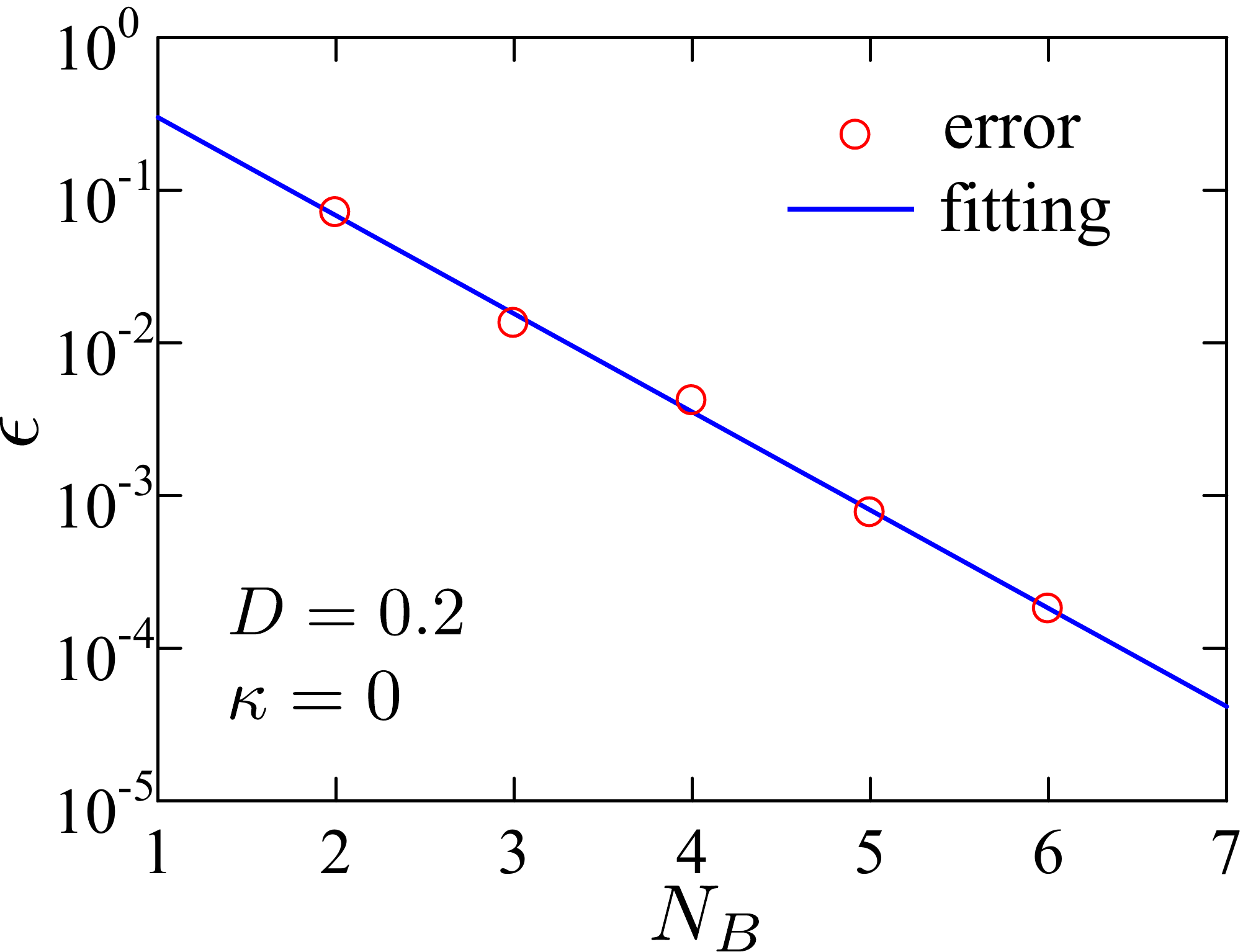}
\par\end{centering}

\caption{Error $\epsilon$ {[}Eq.~\eqref{eq:distance_def}{]} of VSGA as a
function of $N_{B}$ for the one-dimensional stationary case, where
circles and a line denote $\epsilon$ and its fitting curve, respectively
(the curve is $\log\epsilon=-1.48N_{B}+0.28$). Parameters are $D=0.2$
and $\kappa=0$. \label{fig:distance_as_NB}}
\end{figure}

We first study a stationary case (i.e., $I(t)=\kappa$, where $\kappa$
is a constant parameter), because stationary PDFs can be obtained
analytically for a quartic potential. Note that the VSGA in a stationary
case is essentially equivalent to that given in Ref.~\cite{Er:1998:MultiGaussian}.
The stationary PDF $P_{st}(y)$ is given by 
\begin{equation}
P_{st}(y)=\frac{1}{Z(D)}\exp\left[-\frac{U(y)}{D}\right],\label{eq:statdist_1d}
\end{equation}
where $Z(D)=\int_{-\infty}^{\infty}\exp(-U(y)/D)dy$ (numerically
integrated) and $U(x)$ is a potential function $U(y)=-\int\left(y-y^{3}+\kappa\right)dy=y^{4}/4-y^{2}/2-\kappa y$.
The stationary PDF of a VSGA is obtained by letting the system evolve
for a long enough time when it equilibrates. Although in the VSGA,
calculations with larger $N_{B}$ can yield more accurate results,
we employed $N_{B}=5$ (total parameter size is $K=15$ {[}Eq.~\eqref{eq:K_def}{]})
because numerical instability occurs for excessively large $N_{B}$
due to the nonorthogonality of multiple Gaussian distributions. Figure~\ref{fig:statdist_1d}
shows the stationary distributions of the VSGA ($N_{B}=5$; dashed
line) and the analytic solutions obtained for Eq.~\eqref{eq:statdist_1d}
(solid line) for (a) $D=0.2$ and $\kappa=0$; (b) $D=1.0$ and $\kappa=0$;
and (c) $D=0.5$ and $\kappa=0.1$. In all parameter settings, the
VSGA shows very good agreement with the analytical solutions, including
the asymmetric case {[}Fig.~\ref{fig:statdist_1d}(c){]}. In Fig.~\ref{fig:statdist_1d}(a)--(c),
the dot-dashed lines denote the Gaussian bases constituting the PDFs
of the VSGA; we can see that two bases each are located near the deterministic
stable steady states ($y=\pm1$) and one near the deterministic unstable
steady state ($y=0$). In Fig.~\ref{fig:statdist_1d}(a)--(c),
the insets describe tails of PDFs with lot-plots where we see that
tails of VSGA decay slightly slower than analytical solutions; this
is because tails of VSGA are $\exp(-O(y^{2}))$ (Gaussian) while the
analytical ones are $\exp(-O(y^{4}))$. There is a small deviation
in VSGA solutions around $y=0$ and it is considered that the deviation
compensates slow decay at tails of the PDFs. We next see a relation
between $N_{B}$ and an error $\epsilon$ of the approximation by
calculating distance between analytic and VSGA stationary PDFs:
\begin{equation}
\epsilon=\int_{-\infty}^{\infty}\left\{ P_{st}(y)-P(y;\boldsymbol{\theta})\right\} ^{2}dy.\label{eq:distance_def}
\end{equation}
Figure~\ref{fig:distance_as_NB} shows the error $\epsilon$ as a
function of $N_{B}$ with a log-plot, where circles and a line denote
$\epsilon$ and its fitting line ($\log\epsilon=-1.48N_{B}+0.28$),
respectively. We see that the error $\epsilon$ decreases exponentially
as a function of $N_{B}$. 

\begin{figure}
\includegraphics[width=16cm]{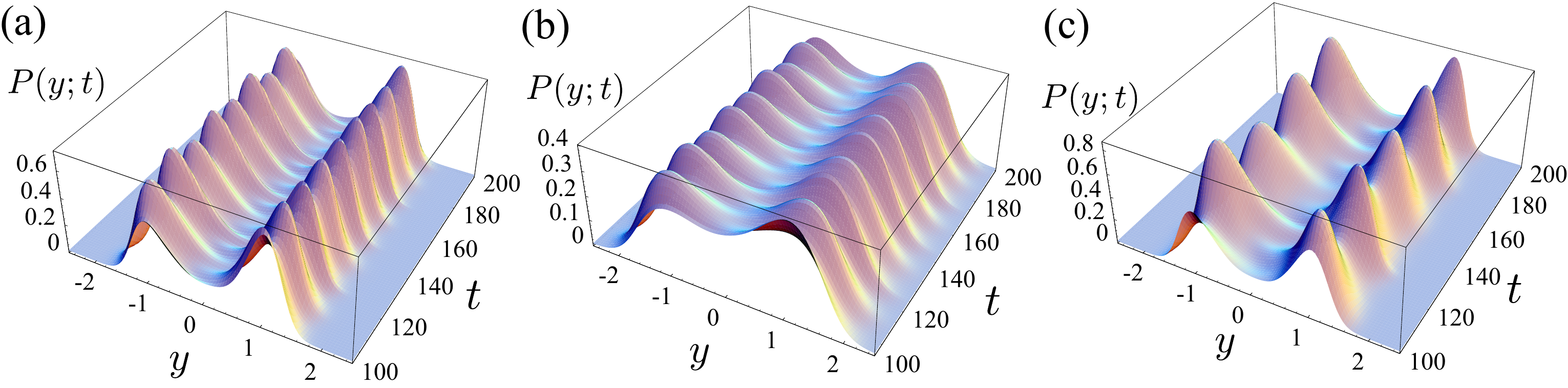}

\caption{(Color online) Dynamical PDFs $P(y;t)$ as functions of $y$ and $t$
obtained by the VSGA with $N_{B}=5$ for (a) $D=0.2$, $\alpha=0.02$
and $\omega=0.5$; (b) $D=1.0$, $\alpha=0.02$ and $\omega=0.5$;
and (c) $D=0.2$, $\alpha=0.02$ and $\omega=0.25$. \label{fig:All3DPlot}}
\end{figure}

\begin{figure}
\includegraphics[width=16cm]{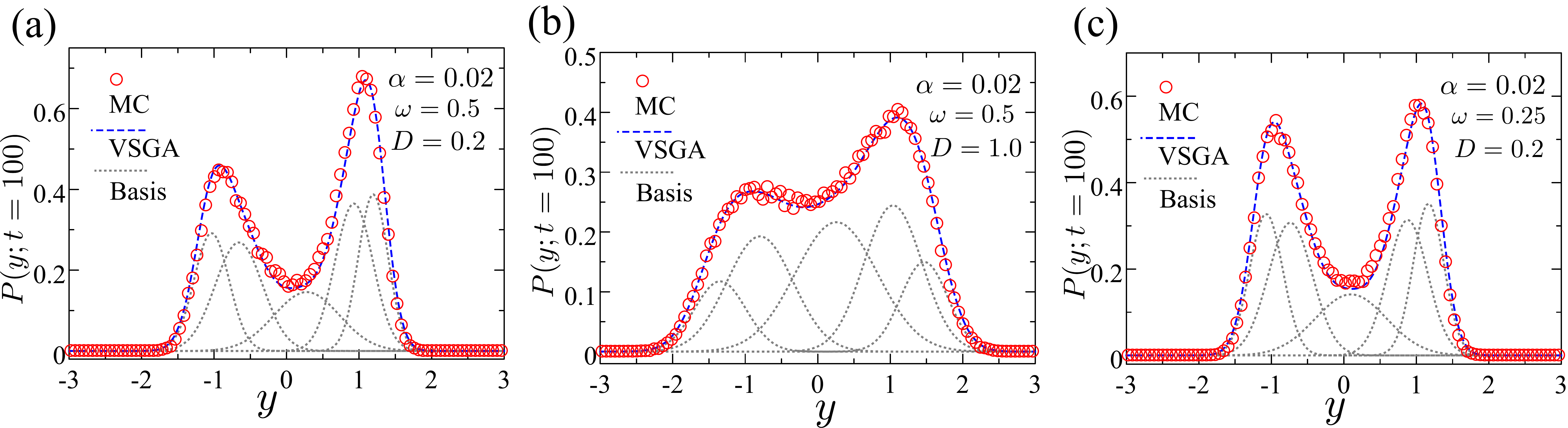}

\caption{(Color online) PDFs $P(y;t)$ at time $t=100$ obtained by MC simulations
(circles), the VSGA with $N_{B}=5$ (dashed line), and each of the
Gaussian bases of the VSGA (dotted line); the parameters in panels
(a)--(c) are the same as in Figs.~\ref{fig:All3DPlot}(a)--(c), respectively.
\label{fig:traj_PDF}}
\end{figure}

We next studied the dynamical case where the input $I(t)$ is given
by Eq.~\eqref{eq:input_definition}. Figure~\ref{fig:All3DPlot}
displays the $P(y;t)$ as functions of $y$ and $t$, which are calculated
by the VSGA with $N_{B}=5$, for (a) $D=0.2$, $\alpha=0.02$ and
$\omega=0.5$; (b) $D=1.0$, $\alpha=0.02$ and $\omega=0.5$; and
(c) $D=0.2$, $\alpha=0.02$ and $\omega=0.25$. To verify the $P(y;t)$
calculated by the VSGA, we evaluated the accuracy of the PDFs $P(y;t)$
at time $t=100$ by calculating the VSGA and by performing MC simulations
(we selected $t=100$ so that we could ignore the effects of the initial
values). For the MC simulations, the PDFs were constructed by repeating
the stochastic simulations $100,000$ times (time resolution is $0.0001$).
Figures~\ref{fig:traj_PDF}(a)--(c) show the PDFs calculated by the
MC simulations (circles), the VSGA (dashed line), and each of the
bases of the VSGA (dotted line). The parameter settings for panels
(a), (b), and (c) correspond to those in Figs.~\ref{fig:All3DPlot}(a),
(b), and (c), respectively. For all parameter settings, the PDFs of
the VSGA are in excellent agreement with those obtained by the MC
simulations; this verifies the reliability of the VSGA with respect
to the PDFs at a specified time.

\begin{figure}
\includegraphics[width=16cm]{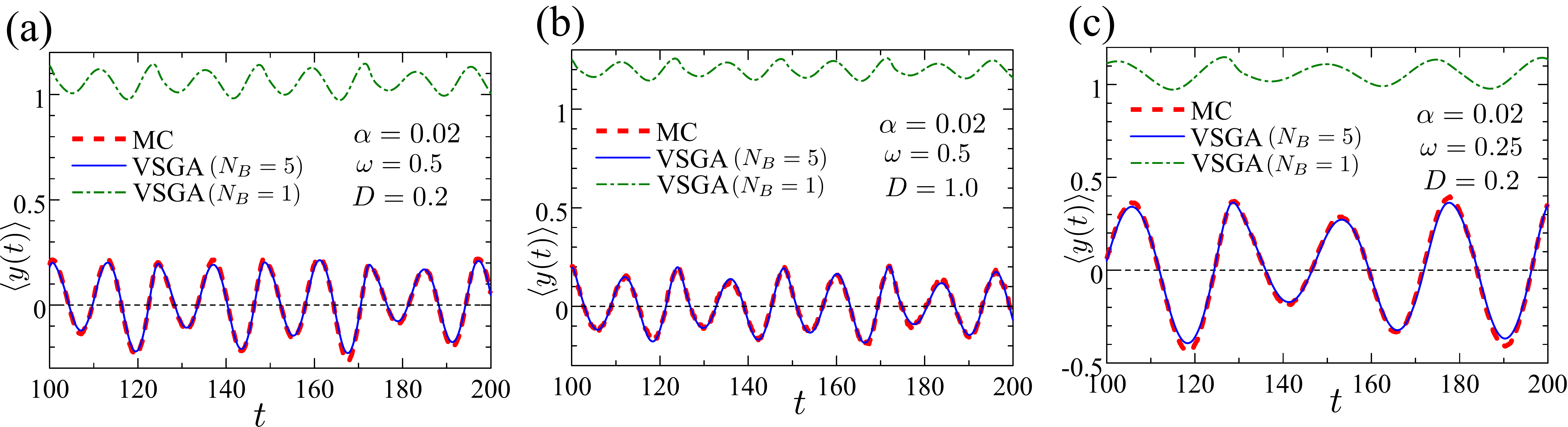}

\caption{(Color online) Mean $\langle y(t)\rangle$ as a function of $t$ as
obtained by MC simulations (dashed line), the VSGA with $N_{B}=5$
(solid line), and the VSGA with $N_{B}=1$ (dot-dashed line); the
parameters in panels (a)--(c) are the same as in Figs.~\ref{fig:All3DPlot}(a)--(c),
respectively. \label{fig:traj_PDF_dynamic}}
\end{figure}

In order to see the dynamical aspects of the VSGA, we also compared
the mean $\left\langle y(t)\right\rangle $ obtained by the MC simulations
to that obtained by the VSGA for the interval $t=100-200$ (we did
not consider the interval $t=0$--$100$ because of the initial value
effects). In Figs.~\ref{fig:traj_PDF_dynamic}(a)--(c), we show the
mean $\left\langle y(t)\right\rangle $ calculated by the MC simulations
(dashed line), the VSGA with $N_{B}=5$ (solid line), and the VSGA
with $N_{B}=1$ (dot-dashed line). The parameter settings for (a),
(b), and (c) correspond to those used in Figs.~\ref{fig:All3DPlot}(a),
(b), and (c), respectively. For the MC simulations, we repeated the
Langevin equations with the same chaotic signal $10,000$ times to
calculate the average. Along with results of VSGA with $N_{B}=5$
plotted by solid lines, dot-dashed lines show those calculated by
the VSGA with $N_{B}=1$ which is similar to the MM case. In Figs.~\ref{fig:traj_PDF_dynamic}(a)--(c),
we can see that the mean $\left\langle y(t)\right\rangle $ of the
values obtained by the VSGA with $N_{B}=5$ are in excellent agreement
with that of the MC simulations, but not with that of the $N_{B}=1$
model. The mean of the $N_{B}=1$ values is located near $1$ and
it only approximates one of the two wells (the mean would be distributed
around $-1$ for particular different initial values).

\begin{figure}
\includegraphics[width=16cm]{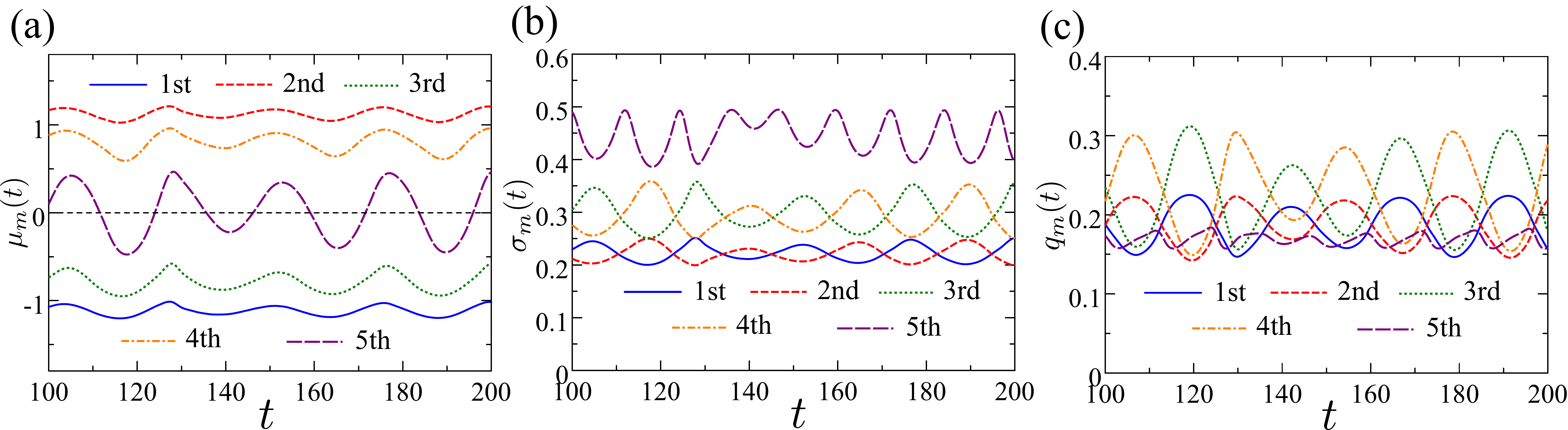}

\caption{(Color online) Time evolution of (a) the mean $\mu_{m}$, (b) the
standard deviation $\sigma_{m}$, and (c) the weight $q_{m}$, for
each Gaussian basis ($N_{B}=5$). In panels (a)--(c), the solid, dashed,
dotted, dot-dashed, and long-dashed lines represent the quantities
for the $1,2,..,5$th Gaussian bases, respectively. The parameters
are $D=0.2$, $\alpha=0.02$ and $\omega=0.25$. \label{fig:equantity_each_Gaussian}}
\end{figure}

In order to study the properties of the VSGA in more detail, we considered
the time evolution of the parameters of each of the Gaussian bases
for $N_{B}=5$. Figures~\ref{fig:equantity_each_Gaussian}(a), (b),
and (c) show the mean $\mu_{m}$, standard deviation $\sigma_{m}$,
and weight $q_{m}$, respectively, of each Gaussian basis as a function
of time $t$ {[}$\mu_{m}$, $\sigma_{m}$, and $q_{m}$ were calculated
by $a_{m}$, $b_{m}$, and $r_{m}$ with Eqs.~\eqref{eq:Sigmai_def}--\eqref{eq:qi_def}{]};
the parameters were $D=0.2$, $\alpha=0.02$ and $\omega=0.25$ identical
to those used in Fig.~\ref{fig:All3DPlot}(c). In Figs.~\ref{fig:equantity_each_Gaussian}(a)--(c),
solid, dashed, dotted, dot-dashed, and long-dashed lines represent
the quantities of the $1,2,..,5$th Gaussian bases, respectively.
In Fig.~\ref{fig:equantity_each_Gaussian}(a), which shows the time
evolution of the mean $\mu_{m}$, we can see that the two wells are
each approximated by two Gaussian distributions, and the temporal
variation of the mean is at most $\sim1$ (cf. $5$th basis; long-dashed
line). From the time evolution of the standard deviation $\sigma_{m}$
{[}Fig.~\ref{fig:equantity_each_Gaussian}(b){]}, we see that the
temporal variation as a function of time is small (about $\sim0.1$),
although the standard deviation averaged over time is different for
each basis. In Fig.~\ref{fig:equantity_each_Gaussian}(c), all the
weights $q_{m}$ are distributed around $0.2$, which shows that all
of the bases contributed to the PDF. Because the VSGA approximates
the two wells with more than two Gaussian distributions, the mean
properly approximates the exact time evolution.

\begin{figure}
\includegraphics[width=11cm]{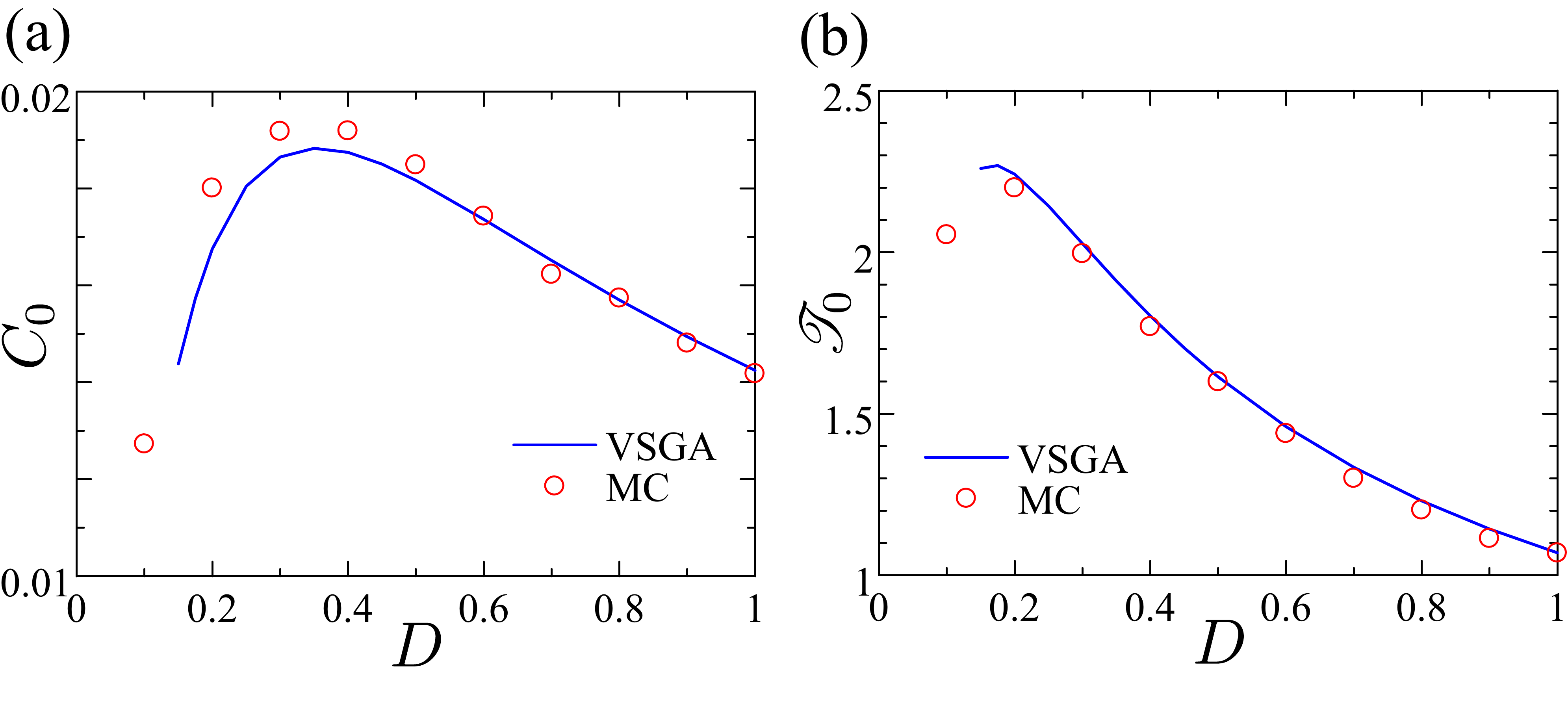}

\caption{(Color online) (a) Correlation $C_{0}$ {[}Eq.~\eqref{eq:ASR_def}{]},
and (b) time-lag $\mathscr{T}_{0}$ {[}Eq.~\eqref{eq:taumax_def}{]}
as a function of the noise intensity $D$, obtained by the VSGA with
$N_{B}=5$ (solid line) and by the MC simulations (circles). The parameters
are $\alpha=0.02$ and $\omega=0.5$. For $D<0.15$, VSGA could not
calculate PDFs. \label{fig:ASR}}
\end{figure}

Driven bistable systems subject to noise are often characterized by
the maximal signal-to-noise ratio under adequate noise strength (stochastic
resonance; SR). Although SR was originally studied in periodic signals,
Ref.~\cite{Collins:1995:ASRinExcite,Collins:1996:AperiodicSR} studied
the SR effects in aperiodic signals. We quantified the extent of SR
for aperiodic signals as follows:
\begin{equation}
C_{0}=\overline{I(t)\left\langle y(t+\mathscr{T}_{0})\right\rangle },\label{eq:ASR_def}
\end{equation}
with 
\[
\overline{\mathcal{F}(t)}=\frac{1}{T}\int_{t_{0}}^{t_{0}+T}\mathcal{F}(t)dt,
\]
where $\mathcal{F}(t)$ is an arbitrary time dependent function, $t_{0}$
is the starting time and $T$ is the duration of the observation;
we again set $t_{0}=100$ and $T=200$. In Eq.~\eqref{eq:ASR_def},
$\mathscr{T}_{0}$ is time lag yielding the maximal correlation: 
\begin{equation}
\mathscr{T}_{0}=\underset{\mathscr{T}}{\mathrm{argmax}\,}\overline{I(t)\left\langle y(t+\mathscr{T})\right\rangle }.\label{eq:taumax_def}
\end{equation}
Here, $C_{0}$ evaluates the amount of chaotic information transmitted,
and a larger value corresponds to better transmission. Although the
SR for a chaotic signal was studied in view of noise-induced phase-synchronization
\cite{Silchenko:1999:ChaosSR}, the ASR obtained by calculating the
correlation of Eq.~\eqref{eq:ASR_def} has not yet been studied.
Figure~\ref{fig:ASR} shows (a) the correlation $C_{0}$, and (b)
the time-lag $\mathscr{T}_{0}$ as a function the noise intensity
$D$, where $C_{0}$ for the VSGA with $N_{B}=5$ is shown by a solid
line and that for the MC simulations is shown by circles. Note that
VSGA could not calculate solutions for $D<0.15$ (see the discussion).
The parameters were $\alpha=0.02$ and $\omega=0.5$. $C_{0}$ achieves
a maximum at $D\simeq0.35$, which indicates the occurrence of ASR.
Comparing the VSGA and MC results shown in Figs.~\ref{fig:ASR}(a)
and (b), we see very good agreement, which verifies the reliability
the VSGA. By using the VSGA, we can calculate properties of chaotically
driven systems without performing stochastic simulations.

\subsection{Chaotically driven bistable potential with colored noise\label{sub:two_dim}}

We next apply the VSGA to a bistable system with colored Gaussian
noise. The one-dimensional colored Gaussian noise system can be embedded
into a two-dimensional Langevin equation with white Gaussian noise:
\begin{align}
\frac{dy}{dt} & =y-y^{3}+I(t)+z(t),\label{eq:color_main}\\
\frac{dz}{dt} & =-\frac{z}{\tau}+\frac{\sqrt{D}}{\tau}\xi(t),\label{eq:color_OU}
\end{align}
where $\xi(t)$ is white Gaussian noise {[}$\left\langle \xi(t)\xi(t^{\prime})\right\rangle =2\delta(t-t^{\prime})${]},
$I(t)$ is the input signal, $\tau$ is the correlation time and $z(t)$
(the Ornstein--Uhlenbeck process) corresponds to a colored noise with
the correlation $\left\langle z(t)z(t^{\prime})\right\rangle =(D/\tau)\exp\left(-|t-t^{\prime}|/\tau\right)$.
We also employed the R\"ossler input for $I(t)$ {[}Eq.~\eqref{eq:input_definition}{]}.
The FPE operator $\hat{L}(y,z,t)$ of Eqs.~\eqref{eq:color_main}--\eqref{eq:color_OU}
is 
\begin{equation}
\hat{L}(y,z,t)=-\frac{\partial}{\partial y}\left(y-y^{3}+z+I(t)\right)+\frac{1}{\tau}\frac{\partial}{\partial z}z+\frac{D}{\tau^{2}}\frac{\partial^{2}}{\partial z^{2}}.\label{eq:color_FPE}
\end{equation}
Substituting Eq.~\eqref{eq:color_FPE} into Eq.~\eqref{eq:MVP_FPE2},
we can again calculate the $K$ (the number of total parameters) coupled
DAE with respect to $\boldsymbol{\theta}=(\boldsymbol{A}_{m},\boldsymbol{b}_{m},r_{m})_{m=1}^{N_{B}}$
with 
\begin{equation}
\boldsymbol{A}_{m}=\left(\begin{array}{cc}
a_{m,11} & a_{m,12}\\
a_{m,21} & a_{m,22}
\end{array}\right),\hspace{1em}\boldsymbol{b}_{m}=\left(\begin{array}{c}
b_{m,1}\\
b_{m,2}
\end{array}\right),\label{eq:A_b_def_2D}
\end{equation}
where $a_{m,12}=a_{m,21}$ ($\boldsymbol{A}_{m}$ is a symmetric matrix).
We require moments of up to the 6th order i.e., $\int_{-\infty}^{\infty}y^{n_{y}}z^{n_{z}}g(y,z;\boldsymbol{A},\boldsymbol{b})dydz$
with $n_{y}+n_{z}\le6$, in order to obtain the DAE. As in the case
with white Gaussian noise, Eq.~\eqref{eq:constraint} should be satisfied
for the normalization, and the resulting $(K+1)$ dimensional DAE
is solved numerically.

\begin{figure}
\includegraphics[width=16cm]{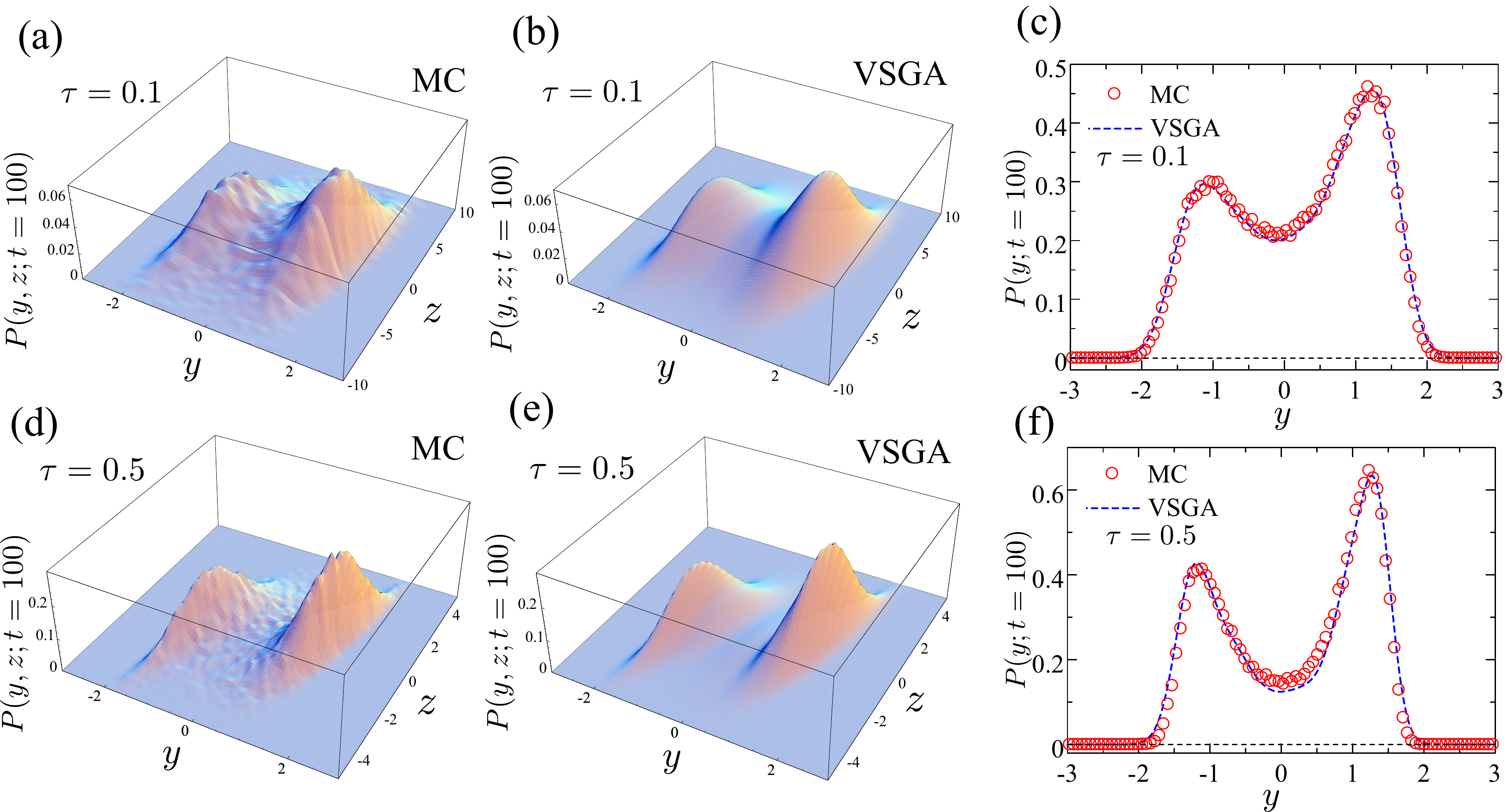}

\caption{(Color online) {[}(a), (b), (d), and (e){]} PDFs $P(y,z,;t)$ at time
$t=100$ for two $\tau$ settings {[}(a) and (b){]} $\tau=0.1$ and
{[}(d) and (e){]} $\tau=0.5$ (other parameters are $D=1.0$, $\alpha=0.02$
and $\omega=0.5$), where (a) and (d) are obtained by the MC simulations
and (b) and (e) are obtained by the VSGA with $N_{B}=5$. {[}(c) and
(f){]} Corresponding marginal PDFs $P(y;t)$ for (c) $\tau=0.1$ and
(f) $\tau=0.5$; results of the VSGA and the MC simulations are shown
by dashed lines and circles, respectively. \label{fig:contour_plots}}
\end{figure}

Figures~\ref{fig:contour_plots}(a) and (b) show PDFs $P(y,z;t)$
at $t=100$, which are calculated by MC simulations and the VSGA with
$N_{B}=5$, respectively, for $\tau=0.1$ with $D=1.0$, $\alpha=0.02$
and $\omega=0.5$. To plot the results of the MC simulations as functions
of $y$ and $z$, we employed kernel distributions. Figure~\ref{fig:contour_plots}(c)
shows the marginal PDF $P(y;t)=\int_{-\infty}^{\infty}P(y,z;t)dz$,
as calculated by MC simulations (circles) and by the VSGA with $N_{B}=5$
(dashed curve). Note that these are in good agreement. Figures~\ref{fig:contour_plots}(d)
and (e) show similar PDFs $P(y,z;t)$, and Fig.~\ref{fig:contour_plots}(f)
shows the marginal PDF $P(y;t)$ for $\tau=0.5$ with $D=1.0$, $\alpha=0.02$
and $\omega=0.5$. As in the one-dimensional case, two bases each
are located near the deterministic stable steady states {[}$(y,z)=(\pm1,0)${]}
and one near the deterministic unstable steady state {[}$(y,z)=(0,0)${]}.
For $\tau=0.5$, the peaks of the PDFs are steeper as can be seen
from Figs.~\ref{fig:contour_plots}(d)--(f). Still the marginal PDF
$P(y;t)$ of the VSGA can approximate the MC simulations. These results
verify the reliability of the VSGA for the systems with colored Gaussian
noise.

\begin{figure}
\includegraphics[width=11cm]{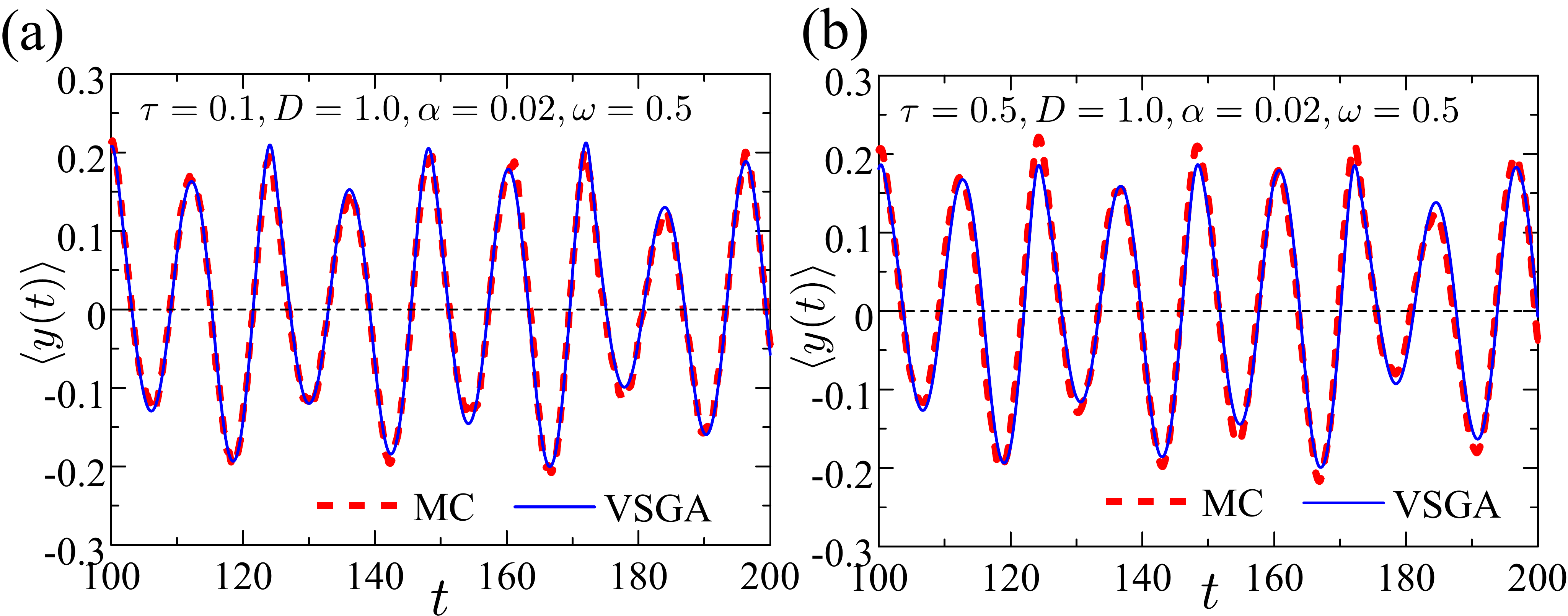}

\caption{(Color online) Time dependence of mean $\langle y(t)\rangle$ for
(a) $\tau=0.1$ and (b) $\tau=0.5$, calculated by MC simulations
(dashed curve) and the VSGA with $N_{B}=5$ (solid curve). The parameters
are $D=1.0$, $\alpha=0.02$ and $\omega=0.5$. \label{fig:moment_color}}
\end{figure}

Next, we evaluated the dynamical properties of the VSGA by comparing
the means $\left\langle y(t)\right\rangle $ of the MC simulations
and the VSGA. Figure~\ref{fig:moment_color} shows the mean $\left\langle y(t)\right\rangle $
calculated by the MC simulation (dashed line) and from the VSGA with
$N_{B}=5$ (solid line), for two different parameters: (a) $\tau=0.1$
and (b) $\tau=0.5$; the other parameters were $D=1.0$, $\alpha=0.02$,
and $\omega=0.5$ (the parameters settings for Figs.~\ref{fig:moment_color}(a)
and (b) correspond to those in Figs.~\ref{fig:contour_plots}(a)--(c)
and (d)--(f), respectively). In Figs.~\ref{fig:moment_color}(a)
and (b), the mean $\left\langle y(t)\right\rangle $ obtained from
the VSGA is in good agreement with that of the MC simulations, for
both $\tau$ values. As in the case with white Gaussian noise, the
VSGA approximates the two wells with more than two Gaussian distributions,
and hence the mean path obtained from the VSGA can accurately approximate
the MC simulations. These results show that the VSGA can be applied
to a two-dimensional system driven by external forces.

\begin{figure}
\includegraphics[width=11cm]{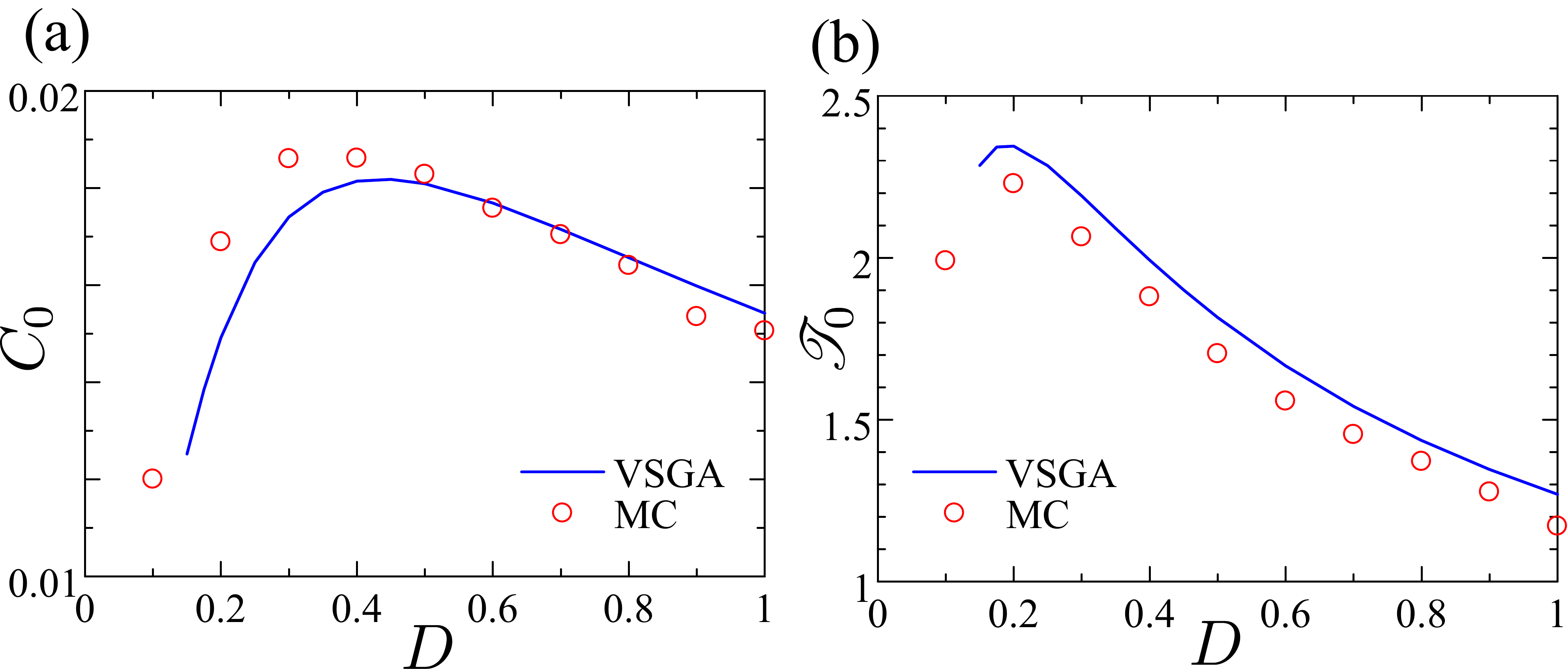}

\caption{(Color online) (a) Correlation $C_{0}$ {[}Eq.~\eqref{eq:ASR_def}{]},
and (b) time-lag $\mathscr{T}_{0}$ {[}Eq.~\eqref{eq:taumax_def}{]}
as a function of the noise intensity $D$, obtained by the VSGA with
$N_{B}=5$ (solid line) and by the MC simulations (circles) for the
colored noise case ($\tau=0.1$). The parameters are $\alpha=0.02$
and $\omega=0.5$. For $D<0.15$, VSGA could not calculate PDFs. \label{fig:C0_color}}
\end{figure}

We also computed the correlation $C_{0}$ {[}Eq.~\eqref{eq:ASR_def}{]}
of a chaotically driven system for the case with colored Gaussian
noise. Figure~\ref{fig:C0_color}(a) shows the correlation $C_{0}$
as a function of the noise intensity $D$ for $\tau=0.1$ where other
parameters are $\alpha=0.02$ and $\omega=0.5$ ($C_{0}$ for the
VSGA with $N_{B}=5$ is shown by a solid line and that for the MC
simulations is shown by circles). As seen in Fig.~\ref{fig:C0_color}(a),
$C_{0}$ also achieved the maximum value at an intermediate value
of $D$. Comparing the VSGA and MC results shown in Fig.~\ref{fig:C0_color}(a),
we see agreement especially for $D>0.4$. Figure~\ref{fig:C0_color}(b)
shows the time-lag $\mathscr{T}_{0}$ {[}Eq.~\eqref{eq:taumax_def}{]}
for the colored-noise case; again $\mathscr{T}_{0}$ for the VSGA
is shown by a solid line and that for MC simulations by circles. From
Fig.~\ref{fig:C0_color}(b), VSGA over-evaluated the time-lag $\mathscr{T}_{0}$,
implying that the reliability of VSGA in the colored-noise case is
worse than the white-noise case. Comparing results of the colored
and white-noise cases, we see that the maximum value of $C_{0}$ for
the colored-noise case is smaller than that for white-noise case which
indicates that the colored noise degrades the ASR effect. However,
when the noise intensity is not optimal (i.e., $D>0.6$), the colored
noise can better transmit information. We note that the time lag for
$\mathscr{T}_{0}$ with the colored-noise case is larger than that
of the white noise.

\section{Discussion and Conclusion\label{sec:discussion}}

The VSGA introduced in Section~\ref{sec:methods} can be used to
obtain several time-dependent solutions of FPEs. We have shown that
our approach can provide very accurate approximations by the superposition
of multiple Gaussian distributions for one- and two-dimensional driven
systems. We have modeled the mean, variance, and weight as time-dependent
parameters. However, as inclusion of the covariance terms {[}$\boldsymbol{A}_{m}$
in Eq.~\eqref{eq:MGMM_def}{]} significantly increases the number
of parameters, it is one possible approach to approximate the covariance
as a constant in order to reduce the computational cost. This approach
has been considered in the Gaussian wavepacket approximation in quantum
mechanics \cite{Skodje:1984:MultiPackets} and is referred to as a
\emph{frozen} method (the time-dependent covariance model is called
a \emph{thawed} method). As shown in Fig.~\ref{fig:equantity_each_Gaussian},
the temporal variation of the standard deviation is around $\sim0.1$
which is smaller than that of the mean. Therefore, if we can first
specify the standard deviation of each Gaussian basis, we may ignore
the time evolution of the variance. This frozen approximation could
dramatically reduce the number of parameters in the VSGA where the
parameter size $K$ is 
\begin{equation}
K=N_{B}(N+1),\label{eq:K_def2}
\end{equation}
being linear with respect to the dimension $N$ and its order of $N$
is smaller than that given by Eq.~\eqref{eq:K_def}. 

Although the effectiveness of the VSGA was demonstrated by our numerical
results, it has some disadvantages. Because multiple Gaussian distributions
are not orthogonal, the VSGA cannot calculate solutions when more
than two Gaussian distributions coalesce. In theory, the accuracy
of the VSGA increases when more basis functions are used (cf. Fig.~\ref{fig:distance_as_NB}).
However, due to their nonorthogonality, an excessively large number
of bases prevents the calculation of the time evolution of the parameters.
Also it becomes more difficult to find valid initial values of DAEs
for the large $N_{B}$ cases. Indeed, for the one-dimensional case,
the VSGA could not calculate the time evolution with $N_{B}=5$ for
$D<0.15$. For such cases, we should reduce $N_{B}$ in order to enable
the calculations. Similarly, for the two-dimensional case, colored
noise with a larger time correlation tends to yield steeper peaks,
which makes the application of the VSGA difficult with large $N_{B}$.
Because peaks of PDFs are represented by a few Gaussian bases in
VSGA, the required number of bases $N_{B}$ can be estimated by the
number of stable points of the system. When there is no input signal,
we can know the number of stable points from a deterministic equation
by solving $f_{i}(\boldsymbol{x})=0$ {[}a time-independent drift
in Eq.~\eqref{eq:Langevin}{]} and evaluating eigenvalues of the
Jacobian matrix around the solutions. If the input signal is weak,
it is expected that a driven case has the same number of stable points
as the no input case. Still the VSGA can provide a computationally
efficient way to calculate the time-dependent dynamics of FPEs. Chaotically
driven stochastic systems have often been solved by MC simulations.
As shown in Section~\ref{sec:results}, the VSGA successfully and
very accurately calculated many of the quantities for the system without
relying on stochastic approaches. 

We applied the VSGA to a quartic bistable potential, where the moments
of the Gaussian can be calculated in closed form. The integral in
Eq.~\eqref{eq:MVP_FPE2} can be computed analytically if the potentials
are represented by polynomials. However, for general nonlinear models,
the moment cannot necessarily be represented in closed form. In such
situations, we may approximate the drift term $f_{i}(\boldsymbol{x})$
by the Taylor expansion: 
\begin{equation}
f_{i}(\boldsymbol{x})\simeq f_{i}(\boldsymbol{\mu}_{g})+\nabla f_{i}(\boldsymbol{\mu}_{g})^{\top}(\boldsymbol{x}-\boldsymbol{\mu}_{g}),\label{eq:f_lin_approx}
\end{equation}
where $\boldsymbol{\mu}_{g}$ is the center of the Gaussian distribution
in the integrand. When using the linear approximation, obtained results
become unreliable when the nonlinearity of a system is strong and/or
the variance of basis is large. As for the quartic bistable case,
the linear approximation can yield accurate solutions when the noise
intensity is sufficiently weak (i.e. the variance of basis is small). 

To summarize, we have proposed the VSGA for the time-dependent solution
for Langevin equations by using the variational principle for superposition
of multiple Gaussian distributions. Because we have shown the effectiveness
of the VSGA in strongly nonlinear systems, the VSGA is expected to
be used for many real-world problems. Applications of the VSGA to
other problems, such as to stochastic models of gene expression \cite{Hasegawa:2013:OptimalPRC,Hasegawa:2014:PRL},
are left to our future study.

\appendix

\section{Relation to conventional multivariate Gaussian representation\label{sec:CG2GR}}

The $N$-dimensional multivariate Gaussian distribution is generally
given by the following representation: 
\begin{equation}
\mathcal{N}(\boldsymbol{x};\boldsymbol{\mu},\boldsymbol{\Sigma})=\frac{1}{(2\pi)^{N/2}\sqrt{|\boldsymbol{\Sigma}|}}\exp\left\{ -\frac{1}{2}(\boldsymbol{x}-\boldsymbol{\mu})^{\top}\boldsymbol{\Sigma}^{-1}(\boldsymbol{x}-\boldsymbol{\mu})\right\} ,\label{eq:conv_multinormal}
\end{equation}
where $\boldsymbol{\mu}$ is the mean vector (column vector) and $\boldsymbol{\Sigma}$
is the covariance matrix (positive definite). In Eq.~\eqref{eq:conv_multinormal},
$|\boldsymbol{\Sigma}|$ denotes the determinant of $\boldsymbol{\Sigma}$.
The mixture of multivariate Gaussian distributions is given by 
\begin{equation}
P(\boldsymbol{x};\{\boldsymbol{\mu}_{m}\},\{\boldsymbol{\Sigma}_{m}\},\{q_{m}\})=\sum_{m=1}^{N_{B}}q_{m}\mathcal{N}(\boldsymbol{x};\boldsymbol{\mu}_{m},\boldsymbol{\Sigma}_{m}),\label{eq:mixture_of_Gaussian}
\end{equation}
where $q_{m}$ is the weight ($\sum_{m=1}^{N_{B}}q_{m}=1$). This
conventional representation and Eq.~\eqref{eq:MGMM_def} are related
in the following way: 
\begin{align}
\boldsymbol{A}_{m} & =\frac{\boldsymbol{\Sigma}_{m}^{-1}}{2},\label{eq:Ai_def}\\
\boldsymbol{b}_{m} & =\boldsymbol{\Sigma}_{m}^{-1}\boldsymbol{\mu}_{m},\label{eq:bi_def}\\
r_{m} & =\frac{q_{m}}{(2\pi)^{N/2}\sqrt{|\boldsymbol{\Sigma}_{m}|}}\exp\left\{ -\frac{1}{2}\boldsymbol{\mu}_{m}^{\top}\boldsymbol{\Sigma}_{m}^{-1}\boldsymbol{\mu}_{m}\right\} .\label{eq:ri_def}
\end{align}
According to Eq.~\eqref{eq:Ai_def}, $\boldsymbol{A}_{m}$ is positive
definite since it is the inverse of a positive definite matrix ($\Sigma$
is positive definite). The inverse transform of Eqs.~\eqref{eq:Ai_def}--\eqref{eq:ri_def}
is 
\begin{align}
\boldsymbol{\Sigma}_{m} & =\frac{1}{2}\boldsymbol{A}_{m}^{-1},\label{eq:Sigmai_def}\\
\boldsymbol{\mu}_{m} & =\frac{1}{2}\boldsymbol{A}_{m}^{-1}\boldsymbol{b}_{m},\label{eq:mui_def}\\
q_{m} & =\frac{r_{m}\pi^{N/2}}{\sqrt{|\boldsymbol{\boldsymbol{A}}_{m}|}}\exp\left\{ \frac{1}{4}\boldsymbol{b}_{m}^{\top}\boldsymbol{A}_{m}^{-1}\boldsymbol{b}_{m}\right\} .\label{eq:qi_def}
\end{align}
In Eqs.~\eqref{eq:Ai_def}--\eqref{eq:qi_def}, we used the fact
that $\boldsymbol{A}_{m}$ is symmetric.

\section{Initial values of DAE\label{sec:init_DAE}}

One of the difficulties in our approach is to find valid initial values
for the DAEs. Unlike conventional (explicit) ordinary differential
equations, DAEs must satisfy an equality condition, and some parameters
should be determined numerically by that equality (in our implementation,
this was done automatically by \emph{MATHEMATICA} 10). We found that
calculating the equality is difficult in some cases. For larger noise
intensities $D$ (for both the white- and colored-noise cases) and
for smaller correlation times $\tau$ (for the colored-noise case),
it is relatively easy to find valid initial values for the DAE. Therefore,
when finding initial values when $D$ is smaller, we first find valid
initial values with $D=1.0$ (large $D$ value) and then iterate the
calculations, adopting the converged stationary values of the preceding
$D$ values as the initial values used to find the next $D$ value.
It is also possible to adjust the system by making $D$ (or $\tau$)
a time-dependent parameter and assuming that $D$ ($\tau$) decreases
(increases) over time starting from large $D$ (small $\tau$) value.
This time-dependent technique was employed for the $\tau=0.5$ case.

\section*{Acknowledgment}

This work was supported by a Grant-in-Aid for Young Scientists B (Y.H.:
No. 25870171) from Ministry of Education, Culture, Sports, Science,
and Technology (MEXT), Japan.

\end{document}